\documentclass[12pt,a4paper,fleqn]{article}

\usepackage[DIV13]{typearea}
\usepackage{amsmath,amssymb,amsfonts}
\usepackage{graphicx}
\usepackage[footnotesize]{caption2}
\usepackage{cite}
\usepackage{mathrsfs}
\usepackage{bbm}
\usepackage{braket}
\usepackage{cancel}

\newcommand{\CiteSeeSaw}{\cite{Minkowski:1977sc,Yanagida:1980,Glashow:1979vf,Gell-Mann:1980vs,Mohapatra:1980ia}}

\DeclareMathOperator{\im}{Im}
\DeclareMathOperator{\diag}{diag}

\newcommand{\mD}{\ensuremath{m_\text{D}}}
\newcommand{\mR}{\ensuremath{m_\text{R}}}
\newcommand{\nuR}{N}
\newcommand{\rep}[1]{\ensuremath{\underline{#1}}} % Representation
\newcommand{\ChargeC}{\mathcal{C}} % Superscript for charge conjugation

\newcommand{\eV}{\ensuremath{\:\text{eV}}}
\newcommand{\GeV}{\ensuremath{\:\text{GeV}}}
\newcommand{\TeV}{\ensuremath{\:\text{TeV}}}

\newcommand{\Eqref}[1]{Eq.~\eqref{#1}}
\newcommand{\Figref}[1]{Fig.~\ref{#1}}

\allowdisplaybreaks[1]

\begin{document}

\begin{titlepage}

\vspace*{1cm}

\begin{center}
{\Large\bf 
Right-Handed Neutrinos at LHC and the Mechanism of Neutrino Mass
Generation\\
}

\vspace{1cm}
\renewcommand{\thefootnote}{\arabic{footnote}}

\textbf{
J\"orn Kersten\footnote[1]{Email: \texttt{jkersten@ictp.it}}$^{(a)}$
and
Alexei Yu.\ Smirnov\footnote[2]{Email: \texttt{smirnov@ictp.it}}$^{(a,b)}$
}
\\[5mm]
\textit{\small $^{(a)}$
The Abdus Salam ICTP, Strada Costiera 11, 34014 Trieste, Italy\\[2mm]
$^{(b)}$
Institute for Nuclear Research, Russian Academy of Sciences,\\
60th Oct.~Anniversary Prospect 7A, 117 312 Moscow, Russia}
\end{center}

\vspace{1cm}

\begin{abstract}
\noindent
We consider the possibility to detect right-handed neutrinos, which are
mostly sing\-lets of the Standard Model gauge group, at future
accelerators.  Substantial mixing of these neutrinos with the active
neutrinos requires a cancellation of different
contributions to the light neutrino mass matrix at the level of
$10^{-8}$.  We discuss possible symmetries behind this cancellation and
argue that for three right-handed neutrinos
they always lead to conservation of total lepton number.
Light neutrino masses can be generated by small
perturbations violating these symmetries. 
In the most general case, LHC physics and the mechanism of
neutrino mass generation are essentially decoupled; with additional
assumptions, correlations can appear between collider observables and
features of the neutrino mass matrix.
\end{abstract}

\end{titlepage}

%%%%%%%%%%%%%%%%%%%%%%%%%%%%%%%%%%%%%%%%%%%%%%%%%%%%%%%%%%%%%%%%%%%%
\newpage

\section{Introduction}
%%%%%%%%%%%%%%%%%%%%%%%%%%%%%%%%%%%%%%%%%%%%%%%%%%%%%%%

The most appealing and natural mechanism for generating small neutrino
masses 
is the (type-I) seesaw mechanism \CiteSeeSaw.  It relies on the
existence of right-handed (RH) neutrinos that are singlets under the
Standard Model (SM) gauge groups and can therefore have large Majorana
masses.  A direct test of the seesaw mechanism would involve the
detection
of these neutrinos at a collider and the measurement of their Yukawa
couplings with the electroweak doublets.  If the Dirac neutrino masses
are similar to the other fermion masses, the Majorana masses turn out to
be of order $(10^8$\,--\,$10^{16})\GeV$,
so that this test is not possible.  In principle, the seesaw mechanism
can also be realised with masses as small as $100\GeV$, though, which
are within the energy reach of the LHC and future colliders.  This
possibility has attracted renewed interest recently, see e.g.\
\cite{delAguila:2005mf,delAguila:2005pf,Bray:2005wv,Han:2006ip,delAguila:2006dx,Atwood:2007fr,Bray:2007ru,deAlmeidaJr.:2007gc,delAguila:2007em}.
However, given the smallness of the light neutrino masses, small RH
masses generically imply tiny Yukawa couplings.  Consequently, also the
mixing between the singlets and the electroweak doublet neutrinos is
tiny, resulting in negligible production cross sections.
In order to allow for large mixing,
different contributions to the light masses have to cancel.  In other
words, the leading-order structure of the mass
matrices leads to vanishing light neutrino masses
\cite{Wyler:1982dd,Bernabeu:1987gr,Buchmuller:1990du,Buchmuller:1991tu,Datta:1991mf,Pilaftsis:1991ug,Ingelman:1993ve,Heusch:1993qu,Tommasini:1995ii,Gluza:2002vs,Pilaftsis:2004xx,Pilaftsis:2005rv,Akhmedov:2006de},
and non-vanishing masses are generated by small perturbations.
Unless this structure can be
motivated by some symmetry, it amounts to fine-tuning.  The known setups
\cite{Wyler:1982dd,Bernabeu:1987gr,Tommasini:1995ii,Pilaftsis:2004xx,Pilaftsis:2005rv}
contain lepton number conservation, which ensures that the heavy states
either form Dirac pairs or decouple from the active neutrinos.

An alternative possibility is that the neutral fermions participating in
the seesaw are not singlets of
the SM symmetry group or have some other interactions that can lead to
their production at future colliders, see e.g.\
\cite{Keung:1983uu,Ferrari:2000sp,Gninenko:2006br,Akhmedov:2006de,Hung:2006ap,Bajc:2006ia,Dorsner:2006fx,Bajc:2007zf,Graesser:2007yj}.
For instance, if all singlets are relatively light, one can
expect that the scale of left-right symmetry is also low.  In this case,
they have gauge interactions with the $W_\text{R}$ and $Z'$.  Then the
discovery is possible for masses up to a few TeV\@.
Another example is the type-III
seesaw mechanism \cite{Foot:1988aq,Ma:1998dn}, where the heavy neutrinos
enter an SU(2) triplet and
therefore can be produced by the electroweak interactions even if their
mixing with light neutrinos is extremely small.

In this paper we will reconsider from the theoretical perspective the
possibility of testing the existence of RH neutrinos at future
colliders.  We study implications of such a detection for the
mechanism of neutrino mass generation.
After discussing the generic estimates that lead to the
expectation of tiny doublet-singlet mixings,
we will consider the cancellation of contributions to light
neutrino masses required by large mixings and possible underlying
symmetries in Sec.~\ref{sec:Mixing}.
Besides the well-known case of lepton number conservation, we will
discuss a scenario based on the discrete symmetry $A_4$ which achieves
the same objective in a different way, but ultimately turns out to
contain a conserved lepton number, too.  We argue that this is a general
feature of any symmetry behind the cancellation.
In Sec.~\ref{sec:Perturbations},
we will systematically study small perturbations of the
leading-order mass matrices that yield viable masses for the light
neutrinos.  In Sec.~\ref{sec:ColliderSignatures}, we will discuss
consequences for signatures at colliders.
Within the setups relying on a symmetry, lepton number violation is
unobservable.  Lepton-flavour-violating processes can have sizable
amplitudes but are difficult to observe at LHC \cite{delAguila:2007em}. 
Consequently, the discovery of RH neutrinos will probably require a
more advanced machine like the ILC.

\section{Cancellations and Symmetries} \label{sec:Mixing}
\subsection{Mixing of Doublet and Singlet Neutrinos}
%%%%%%%%%%%%%%%%%%%%%%%%%%%%%%%%%%%%%%%%%%%%%%%%%%%%%%%%%%%%%%%%%%%%%%%%

In the setup we consider, the Lagrangian responsible for neutrino masses
is the same as in the type-I seesaw scenario \CiteSeeSaw{},
\begin{equation} \label{eq:LMassNu}
	\mathscr{L}_\mathrm{Mass}^\nu =
	- \overline{\nu} m_\mathrm{D} \nuR
	- \frac{1}{2} \overline{\nuR^\ChargeC} m_\mathrm{R} \nuR
	+ \text{h.c.} \;.
\end{equation}
Each RH neutrino%
\footnote{We will call any heavy singlet $N$ that has Yukawa couplings
 with the usual (active) neutrinos a RH neutrino.
},
$\nuR_i$,
generates the (rank 1) contribution to the mass matrix of light
neutrinos
\begin{equation}
m_\nu^{(i)} = - \frac{1}{M_i} \vec{m}_i \vec{m}_i^T \;,
\label{single-m}
\end{equation}
where $M_i$ is the mass of $\nuR_i$ and 
$
\vec{m}_i \equiv (m_{ei}, m_{\mu i}, m_{\tau i})^T
$.
Then the Dirac mass matrix, in the basis where $\mR$ is diagonal, is given
by
$
	\mD = ( \vec{m}_1 , \vec{m}_2 , \vec{m}_3 )
$,
and the complete mass matrix of the light neutrinos equals 
\begin{equation}
	m_\nu =
	\sum_i m_\nu^{(i)} =
	-m_\mathrm{D} m_\mathrm{R}^{-1} m_\mathrm{D}^T \;.
\end{equation}
The Dirac mass terms provide the mixing between the light (active) and
heavy (singlet) states, described by the mixing matrix elements
\begin{equation} \label{eq:nuNMixing}
	V_{\alpha i} = \left(\mD\mR^{-1}\right)_{\alpha i} =
	\frac{m_{\alpha i}}{M_i} \quad
	(\alpha = e, \mu, \tau) \;.
\end{equation}
In terms of $V_{\alpha i}$, the elements of the mass matrix in
\Eqref{single-m} can be rewritten as
\begin{equation}
	(m_\nu^{(i)})_{\alpha \beta} = - V_{\alpha i} V_{\beta i} M_i \;.
\label{mass-mix}
\end{equation}
Assuming the absence of cancellations, the experimental limits on the
light neutrino masses imply that each element
is at most of the order $m_\nu \sim 0.1\eV$.  This yields the upper bound
\begin{equation}
	|V_{\alpha i}| \sim \sqrt{\frac{m_\nu}{M_i}} \lesssim
	10^{-6} \left(\frac{100\GeV}{M_i}\right)^{1/2} \;.
\label{mix-bound}
\end{equation}
It can be considered the generic bound on the mixing of any heavy
Majorana lepton with the light neutrinos.   

The limit (\ref{mix-bound}) is much stronger than the direct bound 
for singlets heavier than the $Z$, obtained from
observations like universality of the weak interactions and the $Z$
width \cite{Bergmann:1998rg,Antusch:2006vw},
\begin{equation} \label{eq:MixingBound}
	\sum_i |V_{\alpha i}|^2 \lesssim 0.01 \;.
\end{equation}
If the heavy neutrinos are to be observable at the LHC or the ILC, their
mixing angles must not lie far below the upper limit \eqref{eq:MixingBound}
\cite{delAguila:2005mf,Han:2006ip,delAguila:2006dx,Bray:2007ru,delAguila:2007em}:
\begin{equation}
	|V_{\alpha i}| \gtrsim 0.01 \;.
\label{theta-lhc}
\end{equation}
Using this value, we obtain from \Eqref{mass-mix} a contribution to the
light neutrino mass
\begin{equation} \label{eq:EstimateMnu}
	m_\nu^{(i)} \sim |V_{\alpha i}|^2 M_i = 
	10^7\eV \left(\frac{|V_{\alpha i}|}{0.01}\right)^2
	 \left(\frac{M_i}{100\GeV}\right). 
\end{equation}
Thus, to reconcile $m_\nu \sim 0.1\eV$ with the
observability of RH neutrinos at the LHC or the ILC, one needs to
arrange a cancellation between the contribution from a given RH neutrino
and some other contribution at the level of
$10^{-8}$.  The situation improves only
slightly if one considers more advanced machines like CLIC or an
$e\gamma$ collider, which could increase the reach in the mixing angle
by about an order of magnitude compared to \Eqref{theta-lhc}
\cite{delAguila:2005pf,Bray:2005wv,delAguila:2006dx}.

In what follows we will discuss cancellations
between the contributions from different RH neutrinos, i.e.\ we
will stay within the framework of the type-I seesaw scenario.  One
could also consider a cancellation with contributions from other
mechanisms, for example involving a Higgs triplet (type-II seesaw
\cite{Lazarides:1981nt,Mohapatra:1981yp,Wetterich:1981bx,Ma:1998dx}),
a fermion triplet (type-III seesaw \cite{Foot:1988aq,Ma:1998dn}) or
a radiatively generated neutrino mass
\cite{Zee:1980ai,Babu:1988ki}.  However, in these cases 
contributions from different, in general unrelated sources have to
cancel, which looks extremely implausible.  The
left-right symmetric models have been suggested as an exception, since
there the type-I and type-II seesaw contributions can be related
\cite{Barenboim:1996vu}.

\subsection{Cancellation of Light Neutrino Masses}
%%%%%%%%%%%%%%%%%%%%%%%%%%%%%%%%%%%%%%%%%%%%%%%%%%%%%%%%%%%%%%%%%

Let us consider first the necessary and sufficient conditions for an
exact cancellation of contributions to the light neutrino masses.
In the case of two RH neutrinos, two matrices have to cancel,
\begin{equation}
m_\nu^{(1)} + m_\nu^{(2)} = 0 \;.
\label{cancel-m}
\end{equation}
Together with \Eqref{single-m} this implies 
\cite{Buchmuller:1990du,Datta:1991mf,Pilaftsis:1991ug}
proportionality of the vectors $\vec m_i$,
\begin{equation}
\vec{m}_1 = y_1 \vec{m}_0 \quad,\quad \vec{m}_2 = y_2 \vec{m}_0 \qquad
(\vec{m}_0 \equiv  m \, (1 , \alpha , \beta)^T) \;,
\label{cancel-vec}
\end{equation}
and 
\begin{equation} 
\label{canc-cond}
        \frac{y_1^2}{M_1} + \frac{y_2^2}{M_2} = 0 \; .
\end{equation}
Therefore, the Dirac mass matrix has the form  
\begin{equation} \label{mDir}
        \mD = m \begin{pmatrix}
                y_1 & y_2 \\
                \alpha y_1 & \alpha y_2 \\
                \beta y_1 & \beta y_2 
                \end{pmatrix} \;. 
\end{equation}

This result can be generalised to the case of three neutrinos
\cite{Buchmuller:1991tu,Ingelman:1993ve,Heusch:1993qu}.  The light
neutrino mass matrix vanishes if and only if the Dirac mass matrix has
rank~1,
\begin{equation} 
\label{eq:mDRank1}
	\mD = m \begin{pmatrix}
	        y_1 & y_2 & y_3 \\
	        \alpha y_1 & \alpha y_2 & \alpha y_3 \\
	        \beta y_1 & \beta y_2 & \beta y_3
	        \end{pmatrix} \;,
\end{equation}
and if
\begin{equation} 
\label{eq:CancelCond}
	\frac{y_1^2}{M_1} + \frac{y_2^2}{M_2} + \frac{y_3^2}{M_3} = 0 \;,
\end{equation}
where the mass parameters are defined in the basis where the singlet
mass matrix is diagonal.
That is, the contributions from the three RH neutrinos to $m_\nu$ have
to be equal
up to a normalisation factor in this case as well.  
Under the conditions (\ref{eq:mDRank1},\ref{eq:CancelCond}), the
light neutrino masses vanish exactly, to all orders in 
$\mD \mR^{-1}$.  This can easily be seen by writing down the
$6\times6$ mass matrix $\mathcal{M}$ and verifying that its rank is 3 or
smaller.  Consequently, the same is true for 
$\mathcal{M}^\dagger \mathcal{M}$, implying the existence of at least 3
vanishing mass eigenvalues.
The $\nu N$-mixing relevant for collider physics, as given by
\Eqref{eq:nuNMixing},
is not restricted by the cancellation condition \eqref{eq:CancelCond}
and hence allowed to be large enough to make the detection of RH
neutrinos possible.

In the following, we will show that Eqs.~\eqref{eq:mDRank1} and
\eqref{eq:CancelCond} are also necessary conditions.
Let us consider the case of $k$ RH neutrinos coupled 
with three active neutrinos. (A general consideration of the case with 
an equal number of left- and right-handed neutrinos has been presented in 
\cite{Heusch:1993qu}.) We parametrise
the contribution of the $i$th RH neutrino to the light 
Majorana mass matrix as 
\begin{equation} 
m^{(i)}_\nu = 
\mu_i  \begin{pmatrix}
                 1 & \alpha_i & \beta_i \\
                 \alpha_i & \alpha_i^2 & \alpha_i\beta_i \\
                 \beta_i & \alpha_i\beta_i & \beta_i^2
                 \end{pmatrix}
\quad (i = 1 \dots k) \;.
\end{equation}
Then the 11-, 12- and 22-elements of the condition 
$m_\nu = \sum_i m^{(i)}_\nu = 0$ can be written as 
\begin{equation} 
\sum_{i=1}^k \mu_i = 0 \quad,\quad
\sum_{i=1}^k \alpha_i \mu_i = 0 \quad,\quad
\sum_{i=1}^k \alpha_i^2 \mu_i = 0 \;. 
\label{system}
\end{equation}
Introducing $x_i \equiv \alpha_i/\alpha_1$, and subtracting 
the first equation in (\ref{system}) from the second and third one,
(divided by $\alpha_1$ and $\alpha_1^2$, respectively) 
we obtain 
\begin{equation} 
\sum_{i=2}^k \left(x_i - 1\right) \mu_i = 0 \quad,\quad
\sum_{i=1}^k \left(x_i^2 - 1\right) \mu_i = 0 \;.
\label{system2}
\end{equation}
\Eqref{system2} is a system of linear equations 
for $\mu_i$.  A similar consideration for the
11-, 13- and 33-elements of the condition $m_\nu=0$
leads to the same system of equations with
$x_i \to x'_i \equiv \beta_i/\beta_1$. 

For $k = 2$ the first equation in 
(\ref{system2}) gives $\mu_2(x_2 -1) = 0$ 
with the unique non-trivial solution $x_2 = 1$ 
or $\alpha_1 = \alpha_2$. Then the second equation 
is satisfied automatically. Similarly one finds 
$\beta_1 = \beta_2$, and consequently  
$m^{(1)}_\nu \propto m^{(2)}_\nu$, so that
we recover Eqs.~(\ref{cancel-vec},\ref{canc-cond}).

For $k = 3$ the system 
\begin{equation} 
\left(x_2 - 1\right) \mu_2  +  \left(x_3 - 1\right) \mu_3 = 0
\quad,\quad
\left(x_2^2 - 1\right) \mu_2  +  \left(x_3^2 - 1\right) \mu_3 = 0
\end{equation}
has non-trivial solutions ($\mu_i \neq 0$) 
only if
$\left(x_2-1\right) \left(x_3-1\right) \left(x_2-x_3\right) = 0$ (zero
determinant).
If this condition is satisfied with $x_2\neq1$ or $x_3\neq1$, one
$\mu_i$ is zero and $\mu_k = -\mu_j$ $(k,j \neq i)$ for the two others. 
This implies that one RH neutrino decouples and the problem is reduced
to the case of two RH neutrinos with cancelling contributions,
cf.\ Eqs.~(\ref{canc-cond},\ref{mDir}).
Thus, the only non-trivial case is $x_2 = x_3 = 1$, i.e.\
$\alpha_1 = \alpha_2 = \alpha_3$.  Analogously,
$\beta_1 = \beta_2 = \beta_3$, and consequently 
$m^{(1)}_\nu \propto m^{(2)}_\nu \propto m^{(3)}_\nu$.
Then the definition \eqref{single-m} straightforwardly leads to 
\begin{equation} 
\vec{m}_1 \propto \vec{m}_2 \propto \vec{m}_3 \;,
\end{equation}
which proves that the rank of the Dirac mass matrix must be $1$.
Writing $\mD$ as in \Eqref{eq:mDRank1} and plugging it into the
condition $m_\nu=0$ finally yields \Eqref{eq:CancelCond}.

In the case of $k = 4$ we have two linear equations 
for three variables $\mu_2, \mu_3, \mu_4$ and therefore 
the zero determinant condition does not apply: 
non-trivial solutions appear even if $x_i \neq \pm 1$. 
This means that the Majorana matrices generated by different 
RH neutrinos are not necessarily 
proportional to each other and non-trivial cancellation 
conditions appear.

One interesting (and the most symmetric) example 
is when cancellations occur between 
two pairs of matrices, for instance
\begin{equation} \label{eq:2DPs}
m^{(1)}_\nu = -  m^{(2)}_\nu \quad,\quad
m^{(3)}_\nu = -  m^{(4)}_\nu \;.
\end{equation}
In this case two combinations of the light neutrinos 
couple with RH neutrinos and the latter form 
two heavy Dirac neutrinos.  For $k=6$, all three combinations
of active neutrinos can couple to RH neutrinos.
In what follows, we will concentrate mainly on the case of three RH
neutrinos.

One can also obtain the cancellation condition using the 
Casas-Ibarra parametrisation \cite{Casas:2001sr} 
for the Dirac mass matrix,
\begin{equation}
\mD = U_\text{PMNS} \sqrt{m_\nu} R \sqrt{m_R} \;,
\label{IC}
\end{equation}
where $R$ is an arbitrary orthogonal matrix, $R R^T = \mathbbm{1}$.
This matrix disappears from the seesaw formula and therefore 
does not influence the light neutrino masses. 
On the other hand, $R$ does influence the Dirac mass matrix and
therefore the mixing of the RH neutrinos with the active neutrinos. 
In fact, the elements of $R$ can be arbitrarily large, so that 
according to (\ref{IC}) one can obtain large $\mD$ (and therefore large
mixing) for arbitrarily small $m_\nu$.
We will show an example in the appendix where the limit 
$m_\nu \rightarrow 0$ but $\sqrt{m_\nu} R =$ \emph{const.}\ recovers the
cancellation conditions.

\Eqref{eq:mDRank1} implies that only the combinations
\begin{equation} 
\tilde{\nu} = 
\frac{\nu_e + \alpha^* \nu_\mu + \beta^* \nu_\tau}{\sqrt{1 + |\alpha|^2 + |\beta|^2}}
\quad,\quad
\tilde N = \frac{y_1 N_1 + y_2 N_2 + y_3 N_3}{\sqrt{\sum_i |y_i|^2}}
\end{equation}
of left- and right-handed neutrinos participate in the Yukawa
interactions.
Two other combinations of the active neutrinos decouple and therefore
remain massless.  The mass of $\tilde{\nu}$ is zero because the
contributions from the different RH components in $\tilde N$ cancel. 
In the next section we will elaborate on this cancellation more and give
a different interpretation.

The fact that only one combination of the left-handed (LH) neutrinos 
$\tilde \nu$ and one combination of the RH neutrinos 
$\tilde N$ couple can follow from a flavour symmetry. 
For example, in the basis of LH states that includes $\tilde \nu$ and 
the RH states that includes $\tilde N$, a U(1) symmetry
with a simple charge assignment can lead to a single coupling. 
The cancellation condition (\ref{eq:CancelCond}) that
involves both the Yukawa couplings and the masses does not show a simple
symmetry in the most general case.  Without a symmetry motivation, it is
a fine-tuning condition and in addition unlikely to be stable
against radiative corrections at the required level.  In the following
we will therefore discuss cases relying on symmetries.

\subsection{Cancellation due to Lepton Number Conservation} 
\label{sec:DiracPair}
%%%%%%%%%%%%%%%%%%%%%%%%%%%%%%%%%%%%%%%

Let us derive a symmetry that leads to the cancellation as well as the
additional constraints it implies for the particular realisation.
According to our consideration in the previous section, only one
combination of the active neutrinos has Yukawa interactions with
singlets.  
Therefore, we consider the system of one active neutrino
$\tilde{\nu}$ and two or three singlets.
We require that all singlets have masses at the electroweak scale or
higher or decouple from the system.
Since there is only one light neutrino,
the only mass that it may have is a Majorana mass. 
The Majorana mass is forbidden if we assign 
to $\tilde{\nu}$ a non-zero lepton number, e.g.\ $L(\tilde\nu) = 1$,
and require it to be conserved in the whole system.%
\footnote{More precisely, we impose a global symmetry U$(1)_L$ under which the SM
 particles have a charge $L$ that equals their lepton number.
 Thus, it leads to the same consequences as lepton number
 conservation in the SM, in particular massless neutrinos and vanishing
 amplitudes for $L$-violating processes.
}
(Notice that in the case of two active components in the system they 
could form a light Dirac neutrino and our argument would not work.)

Next, we determine the lepton numbers of the singlets which ensure that
only one combination couples to $\tilde\nu$ and that the singlets are
massive.
We can rewrite the Dirac mass term in \Eqref{eq:LMassNu} as 
$
\tilde m \overline{\tilde \nu} \tilde N
$,
where 
\begin{equation}
	\tilde m \equiv m \, \sqrt{
	 {\textstyle \sum_i} |y_i|^2 \left(1 + |\alpha|^2 + |\beta|^2\right)
	} \;.
\end{equation} 
This mass term implies that  
$\tilde N$ has the lepton number $L(\tilde N) = 1$, and all other RH
neutrinos have $L \neq 1$.

We first consider the case of two RH neutrinos,
denoting by $N'$ the combination 
of RH components that is orthogonal to $\tilde N$. 
Then the only way to generate a mass 
for $N'$ and $\tilde N$ that is consistent with lepton number 
conservation is to prescribe $L(N') = - 1$ 
and to introduce the mass term $M \overline{N^{\prime\ChargeC}} \tilde N$. 
Combining the mass terms,
\begin{equation} \label{eq:CombinedMassTerms}
-\left( \tilde m \overline{\tilde\nu} + M \overline{N^{\prime\ChargeC}}
\right) \tilde N + \text{h.c.} \;,
\end{equation}
we see that $\tilde \nu$ mixes with $N^{\prime\ChargeC}$ to form a
heavy Dirac neutrino together with $\tilde N$.  This Dirac neutrino has
mass $\sqrt{M^2 + \tilde m^2}$, while the
orthogonal combination of $\tilde\nu$ and $N^{\prime\ChargeC}$
is massless.
In this way we have arrived at the symmetry structure used 
previously in order to obtain the cancellation in 
\cite{Wyler:1982dd,Bernabeu:1987gr,Tommasini:1995ii,Pilaftsis:2004xx,Pilaftsis:2005rv}.

In the case of three singlets, there are two 
combinations $N'_1$ and $N'_2$ orthogonal 
to $\tilde N$, and consequently, several possibilities to realise
lepton number conservation and the cancellation appear
\cite{Branco:1988ex}:
\begin{enumerate}
\item $L(N'_1) = - 1$ and $L(N'_2) \neq \pm 1$ (or vice versa). 
In this case
a Dirac particle arises as before, 
whereas $N'_2$ decouples from the system. It can 
have a Majorana mass if $L(N'_2) = 0$.

\item $L(N'_2) = L(N'_1) = - 1$.  
Now both $N'_1$ and  $N'_2$ can couple to $\tilde N$.  Then the
corresponding combination of $N_1^{\prime\ChargeC}$ and
$N_2^{\prime\ChargeC}$ appears in \Eqref{eq:CombinedMassTerms} and
forms a Dirac pair with $\tilde N$.  The
orthogonal combination is massless and decouples.  
\end{enumerate}

Thus, in all cases with two and three singlets 
we arrive at the same conclusion: 
if the cancellation of active neutrino masses is
a consequence of lepton number conservation, this symmetry leads to 
one decoupled singlet and to the existence of a Dirac fermion formed
predominantly by the other two singlets, i.e.\ the symmetry yields the
structure \eqref{eq:CombinedMassTerms}.
Due to symmetry the structure is stable under radiative corrections. 
In the flavour basis for the LH neutrinos, $L(\nu_\alpha)=1$
for all flavours $\alpha$, and the mass matrices read%
\footnote{A prime denotes quantities in a basis where the singlet mass
 matrix is non-diagonal.
}
\begin{equation} \label{eq:MassMatricesDP}
	\mR' = \begin{pmatrix}
	       0 & M & 0 \\
	       M & 0 & 0 \\
	       0 & 0 & M_3
	       \end{pmatrix} \quad , \quad
	\mD' = m \begin{pmatrix}
	         a & 0 & 0 \\
	         b & 0 & 0 \\
	         c & 0 & 0
	         \end{pmatrix} \;.
\end{equation}
If the decoupled singlet has non-zero lepton number, then $M_3=0$.
We will refer to \Eqref{eq:MassMatricesDP} as the cancellation structure
hereafter.

Instead of lepton number conservation, we can use 
some discrete subgroup of U(1)$_L$
to ensure the cancellation.  For example,
invariance under 
$\tilde \nu \rightarrow i \tilde \nu $, 
$\tilde N \rightarrow i \tilde N$,
$N' \rightarrow - i N'$ does the same job. 
In models of this type,
U(1)$_L$ reappears as an accidental symmetry of the 
discussed mass terms, but it may be broken in 
other sectors explicitly to avoid a
massless Majoron \cite{Chikashige:1980qk,Gelmini:1980re}.
In the case of 4 and more RH neutrinos, more than one 
combination of active neutrinos couples with the singlets,
cf.\ \Eqref{eq:2DPs}, and the arguments presented here do not
apply. 

The suppression of the masses of the active neutrinos is not due to 
the seesaw mechanism but due to mixing with additional states and
mismatch between the number of left- and right-handed fields
\cite{Wyler:1982dd}.
So we can conclude that observation of the RH neutrinos 
at LHC and other colliders would imply that
at least those RH neutrinos do not participate in the seesaw 
mechanism.

\subsection{Three Degenerate Singlets and the Discrete Symmetry $\boldsymbol{A_4}$} 
\label{sec:A4}
%%%%%%%%%%%%%%%%%%%%%%%%%%%%%%%%%%%%%%%%%%%%%%%

In general, the cancellation condition does not require the conservation
of lepton number.  In the basis $\tilde \nu, \tilde N, N'_1, N'_2$ the
sufficient condition for the cancellation is that the determinant of the
$N'_1 N'_2$-block of the singlet mass matrix should be zero.  This does
not forbid entries in the singlet mass matrix that violate lepton number.
Hence, one may ask whether symmetries exist which lead to the
cancellation, but not to $L$-conservation.

Let us first assume that such a symmetry produces equal masses for all
singlets, $\mR = M \, \mathbbm{1}$, and realises \Eqref{eq:CancelCond}
in such a way that seemingly all three singlets participate
in cancelling the light neutrino masses,
\begin{equation} \label{eq:CancelCondEqualMi}
	y_1^2 + y_2^2 + y_3^2 = 0 \;.
\end{equation}
We will now show that in this case there always exist a Dirac pair 
of heavy neutrinos and a decoupled singlet. 
That is, in fact the system does realise lepton number conservation. 

From \Eqref{eq:mDRank1} we know that the Dirac mass terms have the form 
\begin{equation}
	-m \overline{\tilde \nu} \left( y_1 N_1 + y_2 N_2 + N_3 \right) +
	\text{h.c.} \;, 
\label{term2}
\end{equation}
where without loss of generality we have set $y_3 = 1$.  
Now the cancellation condition \eqref{eq:CancelCondEqualMi} reads  
$y_1^2 + y_2^2 = -1$. Recall that $y_i$ are complex parameters and that
in general $|y_i|^2 \neq 1$.
Performing an orthogonal transformation $N_i \rightarrow N_{ir}$,
it is straightforward to check%
\footnote{One has to perform a rotation in the 12-plane that leads to a
 real Yukawa coupling of the second RH neutrino and afterwards a
 rotation in the 23-plane that makes the coupling of the third RH
 neutrino vanish.
}
that the mass term (\ref{term2}) can be reduced to
\begin{equation}
	-\kappa \overline{\tilde \nu} \left( i N_{1r} + N_{2r} \right) +
	\text{h.c.} \;,
\label{term3}
\end{equation}
where $\kappa  = m \sqrt{(\im y_1)^2 + (\im y_2)^2}$. 
Thus, the third singlet decouples.  
As the singlet mass matrix is proportional to the unit matrix,
it does not change under the transformation.

Introducing $\tilde N  = (i N_{1r} + N_{2r})/\sqrt{2}$ 
and the orthogonal combination $N'_1  = 
(- i N_{1r} + N_{2r})/\sqrt{2}$, we find that
the Yukawa couplings and the mass term become  
$\sqrt{2}\kappa \bar{\tilde \nu} \tilde N$ and
$M \overline{N_1^{\prime\ChargeC}} \tilde N$, respectively.
Consequently, $\tilde N$ and $N'_1$ form a Dirac pair and we reproduce
precisely the structure \eqref{eq:MassMatricesDP} that corresponds 
to lepton number conservation. 

As an interesting special case,
let us consider the most symmetric scenario where
$|y_1|^2=|y_2|^2=|y_3|^2=1$ or $y_1=1$, $y_2=\omega$, $y_3=\omega^2$,
where $\omega=e^\frac{2\pi i}{3}$.  This scenario can arise from the
discrete flavour symmetry $A_4$.  Suppose that the singlets transform
under the representation \rep{3}, while all the LH neutrinos
transform under \rep{1''} in the notation of \cite{Altarelli:2006ri}.
Then $\mR = M \, \mathbbm{1}$, and a Dirac
mass matrix is obtained from the interactions
\[
	\sum_{i=1}^3 h_i \, \overline\nu_i \left(
	\nuR_1 \phi_1 + \omega \nuR_2 \phi_2 +
	\omega^2 \nuR_3 \phi_3 \right) \;,
\]
where $h_i$ are coupling constants and $\phi$ is a scalar transforming
under \rep{3} with vacuum expectation values (vevs) 
$\braket{\phi_k} \equiv v_k$.  We find
\begin{equation} \label{eq:A4mD}
	\mD = \begin{pmatrix}
	      h_1 v_1 & \omega h_1 v_2 & \omega^2 h_1 v_3 \\
	      h_2 v_1 & \omega h_2 v_2 & \omega^2 h_2 v_3 \\
	      h_3 v_1 & \omega h_3 v_2 & \omega^2 h_3 v_3
	      \end{pmatrix} \;.
\end{equation}
In the notation of \Eqref{eq:mDRank1}, this corresponds to
$m=h_1 v_1$, $y_1=1$, $y_2=\omega v_2/v_1$ and $y_3=\omega^2 v_3/v_1$,
so that vanishing light neutrino masses are obtained for 
\begin{equation} \label{eq:A4Vev}
	v_1=v_2=v_3=v
\end{equation}
(up to phase factors),
which is required in most mass models based on $A_4$.
If one did not assign the LH neutrinos to a one-dimensional
representation, producing a rank-1 Dirac mass matrix would require
tuning or a non-trivial extension of the symmetry.

Transforming the RH fields into $\tilde N = U_\text{mag}^\dagger N$,
where
\begin{equation}
	U_\text{mag} = \frac{1}{\sqrt{3}} \begin{pmatrix}
	               1 & 1 & 1 \\
	               1 & \omega & \omega^2 \\
	               1 & \omega^2 & \omega
	               \end{pmatrix}
\end{equation}
is the magic matrix, we obtain the Dirac term
$\tilde m \overline{\tilde\nu} \tilde N_3$ with
$\tilde m = \sqrt{3} v \sqrt{|h_1|^2+|h_2|^2+|h_3|^2}$, and the mass
matrix of the RH neutrinos $\tilde N$,
\begin{equation}
	\tilde m_\text{R} = U_\text{mag} U_\text{mag} =
	\begin{pmatrix}
	1 & 0 & 0 \\
	0 & 0 & 1 \\
	0 & 1 & 0
	\end{pmatrix} \;.
\end{equation}
That is, $\tilde\nu$ mixes with $\tilde N_2^\ChargeC$ and forms a Dirac
pair with $\tilde N_3$.
The decoupled singlet $\tilde N_1$ retains the mass $M$, while the Dirac
neutrino is slightly heavier.
Thus, lepton
number conservation arises as an accidental global symmetry in this
$A_4$ toy model.

\subsection{Cancellation without Lepton Number Conservation?}

In the above examples, the singlets contributing to the cancellation
mechanism have equal masses.  Due to the imposed symmetry the
cancellation is stable against radiative corrections. 
Let us relax the requirement of equal masses and
consider the renormalisation group evolution of neutrino masses.
Suppose that the singlets $N_1$ and $N_2$ are relevant for the
cancellation and that the condition \eqref{eq:CancelCond} is imposed by a
symmetry 
at the energy scale
$M_2$, i.e.\ $m_\nu^{(1)}(M_2)=-m_\nu^{(2)}(M_2)$.  Below this scale, 
the symmetry is broken.  We
use an effective theory where $N_2$ is integrated out, so that the
contribution $m_\nu^{(2)}$ corresponds to a dimension-5 operator.  In
the SM, the running of this operator differs from that of $m_\nu^{(1)}$
\cite{Antusch:2002rr,Antusch:2005gp}, so that
\begin{align}
	\frac{d}{dt} m_\nu \Big|_{\mu=M_2} &=
	\frac{d}{dt} m_\nu^{(1)} \Big|_{\mu=M_2} +
	 \frac{d}{dt} m_\nu^{(2)} \Big|_{\mu=M_2} =
	\frac{1}{16\pi^2} \left( \alpha_1 \, m_\nu^{(1)}(M_2) + 
	 \alpha_2 \, m_\nu^{(2)}(M_2)\right)
\nonumber\\
{} &=
	-\frac{1}{16\pi^2} \,
	 \bigl(\lambda + \tfrac{3}{2} g^2 + \tfrac{3}{2} g^{\prime2}\bigr)\,
	 m_\nu^{(1)}(M_2) \;,
\end{align}
where $t \equiv \ln(\mu/\mu_0)$ and $\mu$ is the energy scale.  The term
involving the Higgs self-coupling $\lambda$ and the gauge couplings 
$g, g'$ is of order $1$.  Thus, using the estimate 
$m_\nu^{(1)} \sim 0.01\GeV$ from \Eqref{eq:EstimateMnu}, we obtain a
light neutrino mass of
\begin{equation} \label{eq:MnuFromRunning}
	m_\nu(M_1) \sim \frac{d}{dt} m_\nu \Big|_{\mu=M_2} \Delta t \sim
	10^{-4}\GeV \, \ln\frac{M_2}{M_1}
\end{equation}
at $M_1$, which is unacceptable unless $N_1$ and $N_2$ are nearly
degenerate.  Of course, the problem becomes even worse if also the third
singlet contributes to the cancellation, since then there are additional
corrections from the running between $M_2$ and $M_3$.

As the running is due to diagrams with Higgs fields in the loop, our
estimate is not reliable if the Higgs is heavier than $M_2$.
In supersymmetric theories, both $m_\nu^{(1)}$ and $m_\nu^{(2)}$ obey
the same renormalisation group equation due to the non-renormalisation
theorem, so that $m_\nu$ remains zero above the mass scale
$M_\text{SUSY}$ of the superparticles.  However, as long as
$M_1<M_\text{SUSY}$, the SM can be used as an effective theory below
this scale, so that the estimate \eqref{eq:MnuFromRunning} remains valid
if we replace $M_2$ by $M_\text{SUSY}$.  In any case, further changes of
the neutrino mass matrix from threshold corrections 
\cite{Pilaftsis:1991ug,Pierce:1996zz,Chun:1999vb} tend to yield
too large masses as well, although a simple model-independent estimate
is more difficult.

These arguments suggest that the cancellation of light neutrino masses
can only be realised without fine-tuning, if the RH neutrinos
contributing to the cancellation are nearly degenerate in mass.  As we
have seen in the previous sections, this implies the existence of a
symmetry U(1)$_L$ which guarantees the conservation of total lepton
number in the light sector.  Consequently, any more complicated symmetry
leading to vanishing neutrino masses has to contain U(1)$_L$ as a
subgroup or accidental symmetry.

\section{Non-Zero Neutrino Masses from Perturbations}
\label{sec:Perturbations}
%%%%%%%%%%%%%%%%%%%%%%%%%%%%%%%%%%%%%%%%%%%%%%%%%%%%%%%%%%%%%%%%%%

We will now discuss small perturbations of the cancellation structure
introduced in the previous section that lead to non-vanishing light
neutrino masses.  We will identify the simplest cases which result in
viable neutrino masses and mixings.

\subsection{Generic Perturbations} \label{sec:PerturbationsDirac}

In the case of the Dirac pair of Sec.~\ref{sec:DiracPair}, the most
general possibility for perturbing the cancellation structure
\eqref{eq:MassMatricesDP} is
\begin{equation} \label{eq:GeneralPertDirac}
	\mR' = \begin{pmatrix}
	       \epsilon_1 M & M & \epsilon_{13} M \\
	       M & \epsilon_2 M & \epsilon_{23} M \\
	       \epsilon_{13} M & \epsilon_{23} M & M_3
	       \end{pmatrix}
	\quad , \quad
	\mD' = m \begin{pmatrix}
	         a & \delta_a & \epsilon_a \\
	         b & \delta_b & \epsilon_b \\
	         c & \delta_c & \epsilon_c
	     \end{pmatrix}
		 \equiv m \begin{pmatrix} r , r_\delta , r_\epsilon\end{pmatrix}
	\;.
\end{equation}
Considering the entries in the mass matrices as spurions, we can
immediately see which of them are relevant for the light neutrino
masses.  The latter have a lepton number of $+2$, so that they will
receive contributions from combinations of parameters which also have
$L=+2$.  We have $L(\epsilon_2)=L(r_\delta)=2$, so that these
parameters can contribute directly via terms also involving the large
Yukawa couplings $r$.  As $L(\epsilon_{23})=L(r_\epsilon)=1$, these
quantities will appear quadratically (or in combinations of $2$ or more
different small parameters).  If all perturbations are of the same order
of magnitude, these contributions will be sub-leading.  Finally,
$L(\epsilon_1)$ and $L(\epsilon_{13})$ are negative, so that terms
involving these quantities have to contain at least $2$ more small
parameters.  Consequently, they are almost completely irrelevant for
neutrino masses at the tree level.

Explicitly, we obtain
\begin{align}
	m_\nu = -m^2 & \left[
	 (\mR^{\prime -1})_{11} \, r r^T +
	 (\mR^{\prime -1})_{12} \, (r r_\delta^T + r_\delta r^T) +
	 (\mR^{\prime -1})_{13} \, (r r_\epsilon^T + r_\epsilon r^T) + {}
	\right.
\nonumber\\
& \left. {} +
	 (\mR^{\prime -1})_{22} \, r_\delta r_\delta^T +
	 (\mR^{\prime -1})_{23} \, (r_\delta r_\epsilon^T + r_\epsilon r_\delta^T) +
	 (\mR^{\prime -1})_{33} \, r_\epsilon r_\epsilon^T
	\right]
\label{eq:mnuGeneralDP}
\end{align}
with
\[
	\mR^{\prime -1} = \frac{1}{M_3} \begin{pmatrix}
	  -\frac{M_3}{M} \epsilon_2 + \epsilon_{23}^2 &
	  \frac{M_3}{M} + \frac{M_3}{M} \epsilon_1 \epsilon_2 +
	    \epsilon_{13} \epsilon_{23} &
	  -\epsilon_{23} + \epsilon_2 \epsilon_{13}
	  \\
	  \vphantom{\frac{M_3^2}{M_3}}
	  \frac{M_3}{M} + \frac{M_3}{M} \epsilon_1 \epsilon_2 +
	    \epsilon_{13} \epsilon_{23} &
	  -\frac{M_3}{M} \epsilon_1 + \epsilon_{13}^2 &
	  -\epsilon_{13} + \epsilon_1 \epsilon_{23}
	  \\
	  -\epsilon_{23} + \epsilon_2 \epsilon_{13} &
	  -\epsilon_{13} + \epsilon_1 \epsilon_{23} &
	  1 + 2 \frac{M}{M_3} \epsilon_{13} \epsilon_{23}
	\end{pmatrix} + \frac{1}{M} \mathcal{O}(\epsilon^3) \;.
\]
In the following, we will assume that $\max(a,b,c) \sim 1$, $m/M \sim 0.1$,
$M \sim 0.1\TeV$ (as required by observability of $N_i$ at LHC), that all
$\epsilon_i$ in $\mR'$ are of the same order of magnitude, and that no
severe cancellations occur in \Eqref{eq:mnuGeneralDP}.  Then neutrino
masses $m_\nu \sim 0.1\eV$ require each term in square brackets
to be of order $10^{-9}\TeV^{-1}$ or smaller.  Applying this criterion
to the first term and the second one, respectively, we obtain
\begin{align}
	\epsilon_2 &\lesssim 10^{-10} \;,
\\
	\max(\delta_a,\delta_b,\delta_c) &\lesssim 10^{-10} \;.
\end{align}
Considering the last term with $M_3 \sim 1\TeV$ for concreteness, we
obtain
\begin{equation}
	\max(\epsilon_a,\epsilon_b,\epsilon_c) \lesssim 10^{-4.5} \;,
\end{equation}
i.e.\ this term would be negligible if also 
$r_\epsilon \sim r_\delta$.  From these constraints on $\epsilon_2$,
$r_\delta$ and $r_\epsilon$ it follows that all other terms in
\Eqref{eq:mnuGeneralDP} are negligible under our assumptions, so that
\begin{equation} \label{eq:mnuLeadingDP}
	m_\nu \approx \frac{m^2}{M} \left[
	 \epsilon_2 \, r r^T - (r r_\delta^T + r_\delta r^T) \right] -
	\frac{m^2}{M_3} \, r_\epsilon r_\epsilon^T \;.
\end{equation}
For completeness, we also list the constraint
\begin{equation}
	\epsilon_{23} \lesssim 10^{-4.5}
\end{equation}
for $M_3 \sim 1\TeV$, which can be obtained from the first term in
brackets in \Eqref{eq:mnuGeneralDP}.
As mentioned, the remaining $L$-violating parameters are all but
irrelevant for neutrino masses and thus unconstrained at the tree level.
However, they contribute to one-loop threshold corrections to $m_\nu$
\cite{Pilaftsis:1991ug}.  If only $\epsilon_1$ is non-zero, we find
using the result given in \cite{Pilaftsis:2005rv}
\begin{equation}
	\Delta m_\nu \approx
	\epsilon_1 \frac{g^2}{128\pi^2} \frac{m^2}{M}
	f(M,M_H) \: r r^T \;,
\end{equation}
where $f$ is of order $1$ and depends on $M$ and the Higgs mass.
Requiring for simplicity the corrections to be significantly smaller than the
tree-level masses, $\Delta m_\nu \lesssim 0.01\eV$, we find
\begin{equation} \label{eq:BoundEps1}
	\epsilon_1 \lesssim 10^{-8} \;.
\end{equation}
The parameter $\epsilon_{13}$ violates $L$ by one unit and therefore
enters $\Delta m_\nu$ quadratically.  Furthermore, its contribution is
suppressed by $M/M_3$.  Hence, it is less constrained,
\begin{equation} \label{eq:BoundEps13}
	\epsilon_{13} \lesssim 10^{-3.5} \;,
\end{equation}
again for $M_3 \sim 1\TeV$.
All other perturbations yield negligible radiative corrections if
they satisfy the above tree-level limits.

\subsection{Special Cases} \label{sec:SpecialPerturbationsDirac}
\subsubsection*{Only $\boldsymbol{\epsilon_2\neq0}$}
Returning to \Eqref{eq:mnuLeadingDP}, we see that there are obviously
enough free parameters to fit the measured neutrino mass parameters and
to prevent any observable imprint of the cancellation structure.
Let us therefore look at some more constrained cases.
The simplest possibility is that the
dominant contribution comes from the first term in brackets,
\begin{equation}
	m_\nu \approx \epsilon_2 \frac{m^2}{M} (a , b , c) (a , b , c)^T
	\;.
\end{equation}
Notice that this perturbation generates the singular mass matrix of
light neutrinos that is required by data in the first approximation.
The condition $b=c$ leads to a maximal atmospheric mixing angle.  One
also finds
\begin{equation}
	\sin\theta_{13} = \frac{|a|}{\sqrt{|a|^2+|b|^2+|c|^2}} \;,
\end{equation}
so that the experimental $3\sigma$ limit $\sin^2\theta_{13}<0.04$
\cite{Gonzalez-Garcia:2007ib}
translates into $\left|\frac{b}{a}\right| \gtrsim 3.5$ for $|b| \approx |c|$.

\subsubsection*{$\boldsymbol{\epsilon_2\neq0}$, $\boldsymbol{r_\epsilon\neq0}$,
$\boldsymbol{r_\delta=0}$}
Perturbations of the Dirac mass matrix are needed to generate a second
non-vanishing mass and the solar mixing angle, since the rank of the
product $\mD \mR^{-1} \mD^T$ and thus the number of massive neutrinos is
at most as large as the minimum of the rank of $\mD$ and the rank of
$\mR$.  Adding non-vanishing entries in either $r_\epsilon$ or
$r_\delta$ is sufficient.  If only $r_\epsilon \neq 0$, we can have a
scenario 
where two neutrinos are light due to the U$(1)_L$ symmetry while the
third mass is suppressed by the usual seesaw mechanism, i.e.\ large
$M_3$.  If $r_\epsilon \sim 1$, $M_3$ has to be larger than about
$10^{12}\GeV$ here.  A relatively simple choice of parameters leading to
viable neutrino masses and mixings with a normal mass hierarchy is
\begin{align*}
	m&=10\GeV \;,\; M=100\GeV \;,\; M_3=10^{12}\GeV \;,
\\
	\epsilon_2 &= 2.5 \cdot 10^{-11} \;,\;
	a=0 \;,\; b=-c=1 \;,\;
	\epsilon_a=\epsilon_b=\epsilon_c=0.17 \;.
\end{align*}
In \cite{Pilaftsis:2005rv} an alternative situation was studied where
all singlet masses are equal in the leading order, $M_3=M$.  This is enforced by
an SO(3) flavour symmetry, which contains U$(1)_L$ as a subgroup and
also motivates the smallness of the perturbations in the singlet mass
matrix.

\subsubsection*{Only $\boldsymbol{r_\delta\neq0}$}
If $r_\delta$ is not much smaller than $r_\epsilon$,
its contribution to $m_\nu$ will dominate over that from $r_\epsilon$ as
mentioned above.  We find a particularly interesting case by assuming
that the term which involves $\epsilon_2$ is negligible as well%
\footnote{A non-zero $\epsilon_2$ can be absorbed into
 $r_\delta' \equiv r_\delta + \frac{\epsilon_2}{2} r$, so that it does
 not change the discussion.
}.
Then the neutrino mass matrix
\begin{equation} \label{eq:mnuGeneralPert}
	m_\nu \approx
	-\frac{m^2}{M} \, (r r_\delta^T + r_\delta r^T)
\end{equation}
has rank $2$, so that we can obtain a realistic mass spectrum
with a strong hierarchy from the perturbation $r_\delta$ alone.
Corrections from the neglected terms yield a tiny mass for the lightest
state.  In order to verify that this form of $m_\nu$ is indeed
compatible with the known neutrino masses and mixings, we have
determined values of the parameters that lead to tri-bimaximal mixing
\cite{Harrison:1999cf} and mass squared differences within the
experimentally allowed ranges \cite{Gonzalez-Garcia:2007ib}.  In this case
$\theta_{13}=0$, which places rather strong restrictions on the form of
$m_\nu$.  Nevertheless, solutions for $r$ and $r_\delta$ can be found.
The entries $|a|,|b|,|c|$ have to be roughly of the same order, and
similarly $|\delta_a|,|\delta_b|,|\delta_c|$.
One choice leading to an inverted mass hierarchy and a negative CP
parity for one mass eigenstate is%
\footnote{The Dirac masses were chosen a bit smaller here in order to
 satisfy the bound $|\sum_i V_{e i} V_{\mu i}^*|\lesssim 10^{-4}$ from
 the non-observation of the decay $\mu \to e\gamma$
 \cite{Eidelman:2004wy}.
}
\begin{align*}
	m&=2.8\GeV \;,\; M=100\GeV \;,
\\
	a&=1 \;,\; b = c \approx 0.12 \;,\;
	\delta_a \approx 1.0 \cdot 10^{-10} \;,\;
	\delta_b = \delta_c \approx -4.3 \cdot 10^{-10} \;.
\end{align*}
The mass matrix \eqref{eq:mnuGeneralPert} was studied in the context of
leptogenesis in \cite{Raidal:2004vt}.  It was found that a normal mass
hierarchy with $\theta_{13}$ not far below the experimental bound is
most natural, if there are no hierarchies or special relations between
the parameters in $r$ and $r_\delta$.  The branching ratios for the
flavour-violating decays $l_i \to l_j \gamma$ in supersymmetric seesaw
models turned out to be related via the observed neutrino masses and
mixings and of comparable size.

\subsubsection*{Remarks}
Let us conclude the discussion with some comments about variants of the
scenario, which may be useful input for the construction of models
explaining the perturbations.  In order to reduce the number of free
parameters, one could impose an ``$L$ parity'', i.e.\ a $\mathbbm{Z}_2$
symmetry under which all fields with non-zero lepton number change sign.
Then only the perturbations $\epsilon_1$, $\epsilon_2$ and $r_\delta$
are allowed, which violate $L$ by two units.

If the main goal is avoiding tiny parameters instead, one could use
only terms which violate lepton number by one unit, since they
appear quadratically in the light neutrino mass matrix as mentioned.
This means that only couplings of the singlet $\nuR_3$
contribute at the tree level, leaving two active neutrinos massless.  A
second mass can then be generated by radiative corrections if
$\epsilon_{13}$ is sufficiently large.
Alternatively, one could impose the restriction that all perturbations
are related to a single more fundamental parameter $\varepsilon$
violating $L$ by one unit, i.e.\ 
$\epsilon_{23}, r_\epsilon \sim \varepsilon$ and 
$\epsilon_2, r_\delta \sim \varepsilon^2$.  Then the contributions of
all these parameters to the neutrino masses are of similar sizes, if
$M_3$ is not much larger than $M$.  The dependence of 
$(\mR^{\prime -1})_{11}$ on $\epsilon_{23}$ is non-negligible, and the
term proportional to $(\mR^{\prime -1})_{13}$ in \Eqref{eq:mnuGeneralDP}
becomes relevant in general.  For $m=10\GeV$ and
$M_3 \sim 10 \, M \sim 1\TeV$, the value $m_\nu \sim 0.1\eV$ requires 
$\varepsilon \lesssim 10^{-5}$.
Finally, one could invoke a cancellation of the leading-order
contributions due to $\epsilon_2$ and $r_\delta$, which occurs for
$r_\delta= \frac{\epsilon_2}{2} r$ according to \Eqref{eq:mnuLeadingDP},
in order to allow larger values for these parameters.

\subsection{$\boldsymbol{A_4}$ Model}

One may hope to obtain a connection between the leading-order mass
matrices relevant for LHC and the perturbations responsible for non-vanishing
neutrino masses in the case of the $A_4$ toy model discussed in
Sec.~\ref{sec:A4}, postulating that the perturbations leave a subgroup
of $A_4$ unbroken.  The vevs \eqref{eq:A4Vev} break $A_4$ down to a
$\mathbbm{Z}_3$ subgroup, so that e.g.\ radiative corrections will
generate new couplings that are invariant under
$\mathbbm{Z}_3$ but not under $A_4$ \cite{He:2006dk}.  However, we find
that these do
not change the form of the Dirac mass matrix, similarly to what happens
in the models discussed in \cite{He:2006dk,Hirsch:2007kh}.  The form of
the singlet mass matrix does change because it obtains non-vanishing
off-diagonal entries.  As a consequence of $\mathbbm{Z}_3$, these
entries are all equal and therefore the
active neutrinos remain massless.

Consequently, we have to consider additional symmetry breaking.  The
remaining options are the $\mathbbm{Z}_2$ subgroups of $A_4$.  If the
$A_4$-triplet scalar $\chi$ responsible for this breaking coupled to the
neutrinos via renormalisable interactions, new entries would be
generated in every element of the Dirac mass matrix, destroying all
predictivity.  Let us therefore assume that $\chi$ is a SM singlet,
so that it can couple to the neutrinos only via the non-renormalisable
operator
\[
	\sum_{i=1}^3 \kappa_i \, \overline\nu_i \left(
	 \nuR_2 \phi_3 \chi_1 +
	 \omega \nuR_3 \phi_1 \chi_2 +
	 \omega^2 \nuR_1 \phi_2 \chi_3 \right) +
	\kappa'_i \, \overline\nu_i \left(
	 \nuR_3 \phi_2 \chi_1 +
	 \omega \nuR_1 \phi_3 \chi_2 +
	 \omega^2 \nuR_2 \phi_1 \chi_3 \right)
\]
with couplings $\kappa_i$ and $\kappa'_i$ of dimension (mass)$^{-1}$,
and the Yukawa term
\[
	\lambda \left[
	 (\overline{\nuR_2^\ChargeC} \nuR_3 + \overline{\nuR_3^\ChargeC} \nuR_2)
	  \chi_1 +
	 (\overline{\nuR_3^\ChargeC} \nuR_1 + \overline{\nuR_1^\ChargeC} \nuR_3)
	  \chi_2 +
	 (\overline{\nuR_1^\ChargeC} \nuR_2 + \overline{\nuR_2^\ChargeC} \nuR_1)
	  \chi_3 \right] .
\]
For concreteness, we assume that $\chi$ develops the vev
\begin{equation} \label{eq:VevChi}
	\braket{\chi} = (v_\chi,0,0) \;.
\end{equation}
Then the corrections to the mass matrices are
\begin{equation}
	\Delta \mD = v v_\chi \begin{pmatrix}
	  0 & \kappa_1 & \kappa'_1 \\
	  0 & \kappa_2 & \kappa'_2 \\
	  0 & \kappa_3 & \kappa'_3
	\end{pmatrix} 
	\quad , \quad
	\Delta \mR = \lambda v_\chi \begin{pmatrix}
	  0 & 0 & 0 \\
	  0 & 0 & 1 \\
	  0 & 1 & 0
	\end{pmatrix} \;.
\end{equation}
The complete Dirac mass matrix can have rank $3$.  To first
order in $\kappa_i$, $\kappa'_i$ and 
$\epsilon \equiv \frac{\lambda v_\chi}{M}$, the elements of the light
neutrino mass matrix equal
\begin{equation}
	(m_\nu)_{ij} = \frac{v^2}{M} \left(
	 2 h_i h_j \epsilon - h_j \bar\kappa_i - h_i \bar\kappa_j \right)
\end{equation}
with $\bar\kappa_i \equiv \omega v_\chi \,(\kappa_i + \omega\kappa'_i)$.
This is the mass matrix of \Eqref{eq:mnuLeadingDP} for $r_\epsilon=0$,
which is compatible with observations, see Sec.~\ref{sec:SpecialPerturbationsDirac}.  We obtain a strong
mass hierarchy with the lightest neutrino receiving a mass only from
higher-order corrections.  Note that including the above-mentioned
$\mathbbm{Z}_3$-invariant corrections changes only $h_i$ but not
$\epsilon$ at the considered level of accuracy.  If the position of the
non-zero entry in $\braket{\chi}$ is changed compared to
\Eqref{eq:VevChi}, $\epsilon$ and $\bar\kappa_i$ will change by factors
$\omega$ or $\omega^2$, but the form of $m_\nu$ will remain unaltered.

Thus, we have constructed a pattern of symmetry breaking that produces
perturbations leading to a viable light neutrino mass matrix, which we
had found in the previous section by introducing all possible small
perturbations and assuming some of them to dominate.
The smallness of the perturbations in the Dirac
mass matrix can be motivated by the fact that they arise from
non-renormalisable interactions.

\section{Collider Signatures} \label{sec:ColliderSignatures}

\begin{figure}
\hfill \includegraphics{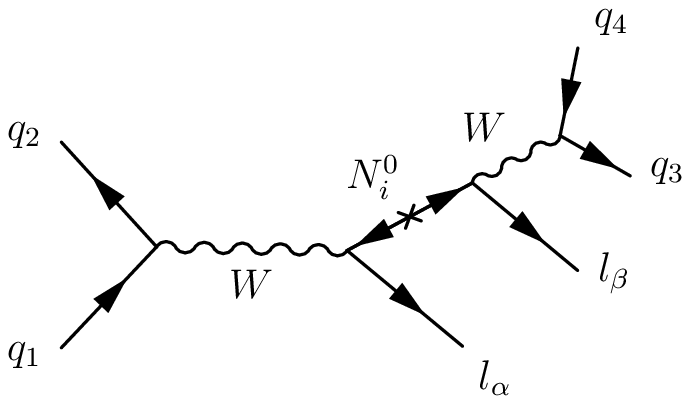} \hfill \includegraphics{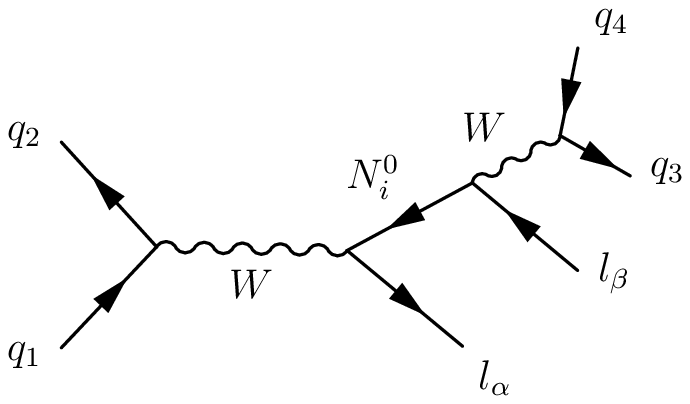} \hfill {}
\caption{Feynman diagrams for lepton-number- (left) and
lepton-flavour-violating processes (right) involving heavy neutrinos at
the LHC.}
\label{fig:Feynmans}
\end{figure}

In this section, we turn to the consequences of the discussed scenarios
for processes involving RH neutrinos at colliders.  Their
charged-current gauge interactions are given by
\begin{equation}
	\mathscr{L}_\text{cc} =
	-\frac{g}{\sqrt{2}} \, \bar{l}_{\alpha} V_{\alpha i} \,
	\gamma^{\mu} W_\mu \frac{1-\gamma_5}{2} \, N^0_i
	+ \text{h.c.} \;,
\end{equation} 
where $l_\alpha$ is a charged lepton and $N^0_i$ is a heavy
neutrino mass eigenstate.  The Feynman diagrams for the most
important processes at LHC \cite{delAguila:2006dx} are shown in
\Figref{fig:Feynmans}.

\subsection{Lepton Number Violation} \label{sec:LNV}

As a promising signal for the production of singlet neutrinos,
$L$-violating processes with like-sign leptons in the final state have
been suggested.  Their amplitudes are proportional to the combination
\begin{equation}
	A_\text{LNV} \equiv V_{\alpha i} \frac{M_i}{p^2-M_i^2+iM_i\Gamma_i}
	V_{\beta i} \;,
\end{equation}
where $\Gamma_i$ is the width of $N^0_i$.
For $M_i \sim 100\GeV$ and $|V_{\alpha i}| \sim 0.1$, one finds
$\Gamma_i \sim 0.01\GeV$ \cite{delAguila:2006dx}.
The quantity $A_\text{LNV}$ also controls the contribution of the RH
neutrinos to neutrinoless double beta decay.  The amplitude is proportional
to $A_\text{LNV}$ with $p^2 \to 0$ and $\alpha=\beta=e$.

All the scenarios for the suppression of the light neutrino masses
discussed above involve the conservation of lepton number, so that
$A_\text{LNV}$ vanishes.  As an explicit example, consider the case of a
heavy Dirac pair.  Then $M_1=M_2=M$, and the mixing matrix of the light and the
heavy neutrinos reads
\begin{equation} \label{eq:NuNMixingDP}
	V \approx \mD \mR^{-1} =
	\frac{m}{\sqrt{2}M} \begin{pmatrix}
	  a \, i & a & 0 \\
	  b \, i & b & 0 \\
	  c \, i & c & 0
	\end{pmatrix}
\end{equation}
in the basis where $\mR$ is diagonal
and in the usual seesaw approximation $m \ll M$.  In order
to check the accuracy of the approximation, we have diagonalised the
$6\times6$ mass matrix $\mathcal{M}$ exactly in the special case
$a=b=c$, finding no significant changes.   Obviously, $A_\text{LNV}$
vanishes for all flavours $\alpha,\beta$.

If $L$ violation is introduced, $A_\text{LNV}$ will be proportional to
the corresponding couplings,
which are restricted to be tiny by the smallness of neutrino masses.
Hence, the suppression of the cross section emerges in a very
similar way as in the usual seesaw scenario.
Sizable lepton number violation would
require the perturbations of the cancellation structure to split the
masses of the singlets forming the Dirac pair by an amount $\Delta M$
significantly
larger than their decay width.  In this case, only one singlet would be
produced on-shell and dominate $A_\text{LNV}$, resulting in a non-zero
amplitude.  If, for instance, $p^2=M_1^2$, then
\begin{equation}
	A_\text{LNV} = 
	\frac{1}{i\Gamma_1} V_{\alpha 1} V_{\beta 1} +
	\frac{M_2}{M_1^2-M_2^2+iM_2\Gamma_2} V_{\alpha 2}V_{\beta 2} \approx
	\frac{1}{i\Gamma_1} V_{\alpha 1} V_{\beta 1} -
	\frac{1}{2 \Delta M} V_{\alpha 2} V_{\beta 2} \;.
\end{equation}
For example, the mass splitting caused by $\epsilon_1$ is roughly 
$\Delta M \approx \epsilon_1 M$.  Consequently, for 
$\Delta M \sim 1\GeV \gg \Gamma_i$, we need $\epsilon_1 \sim 0.01$
(again in the case $M \sim 100\GeV$).  This is still a small perturbation but orders
of magnitude above the bound \eqref{eq:BoundEps1}, so that we cannot
avoid unacceptable active neutrino masses without fine-tuning.
The parameter $\epsilon_{13}$ enters the mass splitting quadratically
and therefore has to be larger than $\epsilon_1$ to achieve the same
splitting $\Delta M \sim 1\GeV$, e.g.\ $\epsilon_{13} \sim 0.3$ for 
$M \sim 100\GeV$ and $M_3 \sim 1\TeV$.  On the other hand, the bound
\eqref{eq:BoundEps13} is weaker than \Eqref{eq:BoundEps1} and
can be further relaxed if one allows the one-loop correction to the
neutrino masses to be of the same order of magnitude as the tree-level
terms.  Leaving aside the problem of explaining the large hierarchy
between the perturbations $\epsilon_i$, lepton number violation via a
large $\epsilon_{13}$ may then be achievable with tuning at the percent
level.

\subsection{Lepton Flavour Violation}

If $L$-violating effects are too small to be observable,
one can still hope to detect events with different lepton flavours such
as $e^-\mu^+$ in
the final state, since these have a relatively small SM background as
well.  According to \cite{delAguila:2007em}, such signals are unlikely
to be observable at LHC, however.  Now the amplitude is proportional to
\begin{equation}
	A_\text{LFV} \equiv
	V_{\alpha i} \frac{\cancel{p}}{p^2-M_i^2} V_{\beta i}^* \;.
\end{equation}
In the considered scenarios, the mechanism leading to the cancellation
of $A_\text{LNV}$ causes the different terms in $A_\text{LFV}$ to add up
constructively.  Again considering the example of
\Eqref{eq:NuNMixingDP}, we obtain
\begin{equation} \label{eq:ALFV}
	A_\text{LFV} = \frac{\cancel{p}}{p^2-M^2} \left(
	 V_{\alpha 1} V_{\beta 1}^* + V_{\alpha 2} V_{\beta 2}^* \right) =
	\frac{\cancel{p}}{p^2-M^2} \frac{m^2}{M^2}
	 (a , b , c)_\alpha (a^* , b^* , c^*)_\beta \;.
\end{equation}
Hence, lepton-flavour-violating (LFV) amplitudes can be sizable.  This also
means that bounds from low-energy searches for rare decays cannot be
avoided by cancellations.  The most stringent limit,
\begin{equation} \label{eq:MuToEGamma}
	\Bigl| \sum_i V_{e i} V_{\mu i}^* \Bigr| = \frac{m^2}{M^2} |a b^*|
	\lesssim 10^{-4} \;,
\end{equation}
comes from the non-observation of the decay $\mu \to e\gamma$
\cite{Eidelman:2004wy}.  In order to have at least a small chance of
observing events at LHC, this condition has to be satisfied with
either $a$ or $b$ being very small and the other parameter of order $1$.
In the most minimal examples for perturbations we have seen that the
large atmospheric mixing angle implies $|b|\sim|c|$.  In this
case, all amplitudes would be suppressed if $b$ were small.  Therefore,
making $a$ tiny is the more favourable option.  Then flavour-violating
processes with electrons in the final state are not observable, leaving
processes with the final state $\tau^\pm \mu^\mp$ as the best candidate
for observing singlets.

At the ILC, the situation is more hopeful, since there the resonant
production of RH neutrinos is possible for $|V|_{ei}\gtrsim0.01$
\cite{delAguila:2005mf,delAguila:2006dx}, which is allowed by
\Eqref{eq:MuToEGamma} even if $|a|\sim|b|$.  By observing the branching
ratios for the subsequent decays into charged leptons, one could
determine the mixings with the different flavours directly.

\subsection{Decoupling of Collider Physics from the Light Masses}

If the observation of RH neutrinos at colliders is to shed light onto
the mechanism of neutrino mass generation, the first key question we
have to ask is whether the perturbations responsible for neutrino masses
could have consequences for signals at colliders.  Unfortunately, the
smallness of the light neutrino masses immediately tells us that the
answer is negative.  All perturbations of the couplings of relatively
light singlets yielding neutrino masses are restricted to be tiny.
Thus, they will not lead to observable collider signatures.  
Instead, collider experiments are only sensitive to the large Yukawa
couplings in \Eqref{eq:MassMatricesDP}, i.e.\ to the cancellation structure of the mass matrices
which does not produce neutrino masses.

This leads to the second key question, whether perturbations can be
introduced in such a manner that the light neutrino mass matrix still
``remembers'' in some way the cancellation structure.  In other words,
can perturbations lead to particular features of the light neutrino mass
matrix, so that the cancellation structure is imprinted in the
structure of $m_\nu$?
As argued above, a light neutrino mass matrix with at least two
non-vanishing eigenvalues can only be obtained if the Dirac mass matrix
is perturbed.  In general, this introduces many new parameters, so that
there is little hope to find a simple connection between $m_\nu$ and the
cancellation structure.  Then the answer to the second
question is negative, too.

The situation is better in constrained setups where only some of
the perturbations are present or dominant.  In the cases we discussed, a
strong mass hierarchy is expected.  The number of
free parameters is large enough to reproduce any mixing pattern, so that
there are no definite predictions for the mixing angles.  However, to the extent that the
leading-order Yukawa couplings are fixed by the measured neutrino masses
and mixings, correlations between the branching ratios of LFV processes
can be obtained, cf.\ \Eqref{eq:ALFV}, analogously to what was found for the branching ratios
of LFV decays \cite{Raidal:2004vt}.
As we have argued, the severe limit from $\mu \to e\gamma$ probably
means that at most one LFV branching ratio will be measurable at LHC\@.
Then the predicted correlations can be falsified by observing a
second LFV process.  In order to verify them, one has to determine the
mixings of RH neutrinos with the different flavours directly, which may
be possible at $e^+e^-$ colliders
\cite{delAguila:2005mf,delAguila:2005pf,delAguila:2006dx}.
In the most optimistic case, collider experiments could even test some
predictions of leptogenesis models
\cite{Raidal:2004vt,Pilaftsis:2005rv}.

\section{Summary and Discussion}

We have critically re-examined the possibility of a direct test of the
seesaw mechanism of neutrino mass generation in collider experiments.
We have assumed the existence of right-handed (RH) neutrinos with
masses close to the electroweak scale (but no other new particles or
interactions).  The upper bound on the
light neutrino masses immediately leads to the conclusion that these RH
neutrinos are much too weakly coupled to the Standard Model particles to
be produced at colliders.  This conclusion can only be avoided, if there
is a strong cancellation between the contributions from different RH
neutrinos to the light neutrino masses, which can be due to lepton
number conservation.  Then the seesaw mechanism itself plays only a
minor role in explaining the smallness of neutrino masses.  Light
neutrino masses can appear as a result of a small breaking of lepton
number.  Other effects of this breaking are too small to be tested at
accelerators.  The only hope to check this possibility
is to discover the RH neutrinos and to establish correlations, which may
exist in special cases, between properties of the light neutrino mass
matrix and processes at accelerators involving the RH neutrinos. 
Of course, one cannot exclude the existence of additional, very heavy
RH neutrinos contributing to neutrino masses via the standard seesaw
mechanism, but this cannot be tested directly.

More explicitly,
an exact cancellation of light neutrino masses occurs, if only one
combination of the active neutrinos, $\tilde\nu$, couples with the RH
neutrinos, while two others decouple and remain massless.  The
contributions of the RH neutrinos to the mass of $\tilde\nu$ cancel due
to a certain correlation between their masses and Yukawa couplings.  We
have shown that these are necessary conditions in scenarios with two and
three RH neutrinos.  If there are more than three RH neutrinos, the
conditions do not apply.  In this case, more than one combination of
active neutrinos can couple to the RH neutrinos, and the cancellation
can be realised in a more complicated way. 

We have discussed examples where the cancellation is due to a
symmetry.  In the simplest setup,
one RH neutrino decouples from the system.  Another one mixes with
$\tilde\nu$ and forms a Dirac pair with the third RH neutrino, and the
combination orthogonal to this mixture stays massless.
This structure implies conservation of lepton number.
We have also presented a simple model based on the discrete symmetry
$A_4$, in which $L$ conservation arises as an accidental symmetry.
  
If the cancellation is realised by a symmetry that leads to
lepton number conservation, it is stable
against radiative corrections: three neutrinos remain massless.  In all
other cases, the cancellation is unstable and therefore requires
fine-tuning in several orders of perturbation theory.  This is true both
for setups without any symmetry motivation and for scenarios relying on
a symmetry which does not imply $L$ conservation.

Light neutrino masses are obtained from small perturbations of the
leading-order mass matrices.  We have systematically considered 
all possible perturbations of the mass matrices arising in the
$L$-conserving setup.  
In the $A_4$ toy model, we have discussed a pattern of
symmetry breaking that leaves a $\mathbbm{Z}_2$ subgroup unbroken,
resulting in a subset of the most general perturbations and partially
motivating their smallness.

Thus, both lepton number violation and active neutrino masses arise due
to small perturbations, and their magnitudes are related.  Therefore, we
expect lepton-number-violating signals at colliders to be unobservable
in untuned scenarios.
The cross sections for lepton-flavour-violating processes are not
suppressed, so that LHC might have a chance to observe such reactions.
If this is the case, lepton flavour violation should also be observable
in upcoming experiments studying the decays of charged leptons.

The flavour pattern of processes with RH neutrinos at colliders
depends on the particular combination $\tilde\nu$.  With a realistic
experimental accuracy, a measurement of the perturbations will not be
possible.  Consequently, in the most general case the theory contains
too many free parameters to realise a simple connection between collider
observables and the masses and mixings of the
active neutrinos.  In this sense, the mechanism of neutrino mass 
generation and collider physics decouple.  However, in minimal cases,
where only some perturbations of the cancellation structure are present,
one can find correlations between features of neutrino masses and
accelerator observables.  These could be falsified at the LHC and
tested at the ILC\@.
Of course, the discovery of Standard Model singlets close to
the electroweak scale would be very interesting by itself in any case.

\section*{Acknowledgements}
We would like to thank Evgeny Akhmedov, Stefan Antusch, Wilfried
Buchm\"uller, Tsedenbaljir Enkhbat and Goran Senjanovi{\'c} for helpful
discussions.

\appendix

\section{Cancellation and Casas-Ibarra Parametrisation}
%%%%%%%%%%%%%%%%%%%%%%%%%%%%%%%%%%%%%%%%%%%%%%%%%%%%%%%%%%%

Let us demonstrate how the cancellation condition can be derived using
the Casas-Ibarra parametrisation \cite{Casas:2001sr}.
For illustration we consider the two-generation case and choose 
\begin{equation}
R = 
\begin{pmatrix}
\cosh x & -i\sinh x\\
i\sinh x & \cosh x
\end{pmatrix} \;.
\end{equation}
For simplicity we neglect neutrino mixing, $U_\text{PMNS}=\mathbbm{1}$.
Then using \Eqref{IC} with the diagonal matrices  
$m_\nu  = \diag(m_{\nu 1}, m_{\nu 2})$ and $\mR = \diag(M_1, M_2)$
we can write 
\begin{equation}
\mD = 
\begin{pmatrix}
\cosh x  \sqrt{m_{\nu 1} M_1 } & -i\sinh x \sqrt{m_{\nu 1} M_2 }  \\
i\sinh x \sqrt{m_{\nu 2} M_1 } & \cosh x \sqrt{m_{\nu 2} M_2 }
\end{pmatrix} \;.
\end{equation}
For $x \gg 1$, $\cosh x \approx \sinh x \approx e^x/2$.
If in the limit $m_\nu \rightarrow 0$ the products
$e^x\sqrt{m_{\nu 1}} \rightarrow \sqrt{\mu} =$ \emph{const.},
and 
$e^x\sqrt{m_{\nu 2}} \rightarrow\sqrt{\mu}\alpha =$ \emph{const.}, 
the Dirac mass matrix becomes
\begin{equation}
\mD = 
\begin{pmatrix}
\sqrt{\mu M_1 } &  - i \sqrt{\mu M_2 }  \\
i \sqrt{\mu M_1 }\alpha &  \sqrt{\mu M_2} \alpha
\end{pmatrix} \;.
\end{equation}
This matrix has rank 1 and satisfies the cancellation condition
\eqref{canc-cond}.

\bibliography{Neutrinos}
\bibliographystyle{NewArXiv}

\end{document}